
%
\input phyzzx
%
%
\catcode`@=11
%
%
\font\fourteenmib=cmmib10 scaled\magstep2   \skewchar\fourteenmib='177
\font\twelvemib=cmmib10 scaled\magstep1     \skewchar\twelvemib='177
\font\elevenmib=cmmib10 scaled\magstephalf  \skewchar\elevenmib='177
\font\tenmib=cmmib10                        \skewchar\tenmib='177
%
\font\fourteenbsy=cmbsy10 scaled\magstep2   \skewchar\fourteenbsy='60
\font\twelvebsy=cmbsy10 scaled\magstep1     \skewchar\twelvebsy='60
\font\elevenbsy=cmbsy10 scaled\magstephalf  \skewchar\elevenbsy='60
\font\tenbsy=cmbsy10                        \skewchar\tenbsy='60
%
%
\newfam\mibfam
%
%
\def\fourteenf@nts{\relax
    \textfont0=\fourteenrm          \scriptfont0=\tenrm
      \scriptscriptfont0=\sevenrm
    \textfont1=\fourteeni           \scriptfont1=\teni
      \scriptscriptfont1=\seveni
    \textfont2=\fourteensy          \scriptfont2=\tensy
      \scriptscriptfont2=\sevensy
    \textfont3=\fourteenex          \scriptfont3=\twelveex
      \scriptscriptfont3=\tenex
    \textfont\itfam=\fourteenit     \scriptfont\itfam=\tenit
    \textfont\slfam=\fourteensl     \scriptfont\slfam=\tensl
    \textfont\bffam=\fourteenbf     \scriptfont\bffam=\tenbf
      \scriptscriptfont\bffam=\sevenbf
    \textfont\ttfam=\fourteentt
    \textfont\cpfam=\fourteencp
    \textfont\mibfam=\fourteenmib   \scriptfont\mibfam=\tenmib
    \scriptscriptfont\mibfam=\tenmib }
\def\twelvef@nts{\relax
    \textfont0=\twelverm          \scriptfont0=\ninerm
      \scriptscriptfont0=\sixrm
    \textfont1=\twelvei           \scriptfont1=\ninei
      \scriptscriptfont1=\sixi
    \textfont2=\twelvesy           \scriptfont2=\ninesy
      \scriptscriptfont2=\sixsy
    \textfont3=\twelveex          \scriptfont3=\tenex
      \scriptscriptfont3=\tenex
    \textfont\itfam=\twelveit     \scriptfont\itfam=\nineit
    \textfont\slfam=\twelvesl     \scriptfont\slfam=\ninesl
    \textfont\bffam=\twelvebf     \scriptfont\bffam=\ninebf
      \scriptscriptfont\bffam=\sixbf
    \textfont\ttfam=\twelvett
    \textfont\cpfam=\twelvecp
    \textfont\mibfam=\twelvemib   \scriptfont\mibfam=\tenmib
    \scriptscriptfont\mibfam=\tenmib }
\def\tenf@nts{\relax
    \textfont0=\tenrm          \scriptfont0=\sevenrm
      \scriptscriptfont0=\fiverm
    \textfont1=\teni           \scriptfont1=\seveni
      \scriptscriptfont1=\fivei
    \textfont2=\tensy          \scriptfont2=\sevensy
      \scriptscriptfont2=\fivesy
    \textfont3=\tenex          \scriptfont3=\tenex
      \scriptscriptfont3=\tenex
    \textfont\itfam=\tenit     \scriptfont\itfam=\seveni  
    \textfont\slfam=\tensl     \scriptfont\slfam=\sevenrm 
    \textfont\bffam=\tenbf     \scriptfont\bffam=\sevenbf
      \scriptscriptfont\bffam=\fivebf
    \textfont\ttfam=\tentt
    \textfont\cpfam=\tencp
    \textfont\mibfam=\tenmib   \scriptfont\mibfam=\tenmib
    \scriptscriptfont\mibfam=\tenmib }
\def\mib{\noexpand\f@m\mibfam}
%
%
\Twelvepoint
\catcode`@=12
%

\def\part{\partial}
\def\e#1#2{\tilde e^#1_{~#2}}
\def\A#1#2{A_#1^{~#2}}
\def\F#1#2{F_{#1}^{~~#2}}
\def\w#1#2{{}^{+)}\omega_{#1}^{~#2}}
\def\s#1{\sigma^#1}
\def\un{N\kern-12pt\lower 5pt\hbox{$\displaystyle\sim $~}}
\def\vec#1{{\mib #1}}
\def\diag{{\rm diag~}}
%
\titlepage
\hfill{TIT/HEP-297/COSMO-56}
\title{
	Gravitational Instantons in Ashtekar's Formalism
}
\vskip 2cm
\centerline {Hideki Ishihara,\quad Hiroto Kubotani\dag}
\vskip 0.5cm
\centerline{and}
\vskip 0.5cm
\centerline {Takeshi Fukuyama\ddag}
\vskip 1cm
\noindent
\address{ Department of Physics, Tokyo Institute of Technology,
 Megro-ku Tokyo 152, Japan}
\address{\dag Department of Physics, Ochanomizu University,
Bunkyo-ku Tokyo 112, Japan}
\address
{ ${}\ddag $ Department of Physics, Ritsumeikan University,
  Kusatsu Shiga 525, Japan}
\vskip 1cm

\abstract{
Gravitational instantons of Bianchi type IX space are constructed
in Ashtekar's canonical formalism. Instead of solving the self-duality
condition, we fully solve the constraint on the \lq\lq initial surface\rq\rq\
and \lq\lq Hamiltonian equations\rq\rq .  This formallism is applicable to
the matter coupled system with cosmological constant.
}

\vskip3cm
\endpage

\chapter{Introduction}

In Ashtekar's canonical formalism of general relativity,\Ref\Ash{
	A.Ashtekar, {\sl Phys. Rev.} {\bf D36}, 1587 (1987).
}
gravity is described as the singular system but with only polynomial
constraints.
This bonus is paid with the expense of extending the canonical variables to
complex ones. So the system is subject to so called reality condition which
gurantees the equivalence with Einstein gravity. This reality condition,
however, may spoil the nice feature of Ashtekar's formalism.\Ref\KF{
 K.Kamimura and T.Fukuyama, {\sl Phys. Rev.} {\bf D41}, 1885 (1990).
}
This condition comes from the fact that the (anti) self-duality
complexizes spin connection  in Lorenzian region.

In a quantum theory of gravity, gravitational instantons\Ref\EGH{
 T.Eguchi, P.B.Gilkey and A.J.Hanson, {\sl Phys. Rep.} {\bf 66C}, 213 (1980).
}
are expected to play essential roles as Yang-Mills instantons do
in quantum chromo-dynamics.\Ref\instanton{
 A.A.Belavin, A.M.Polyakov, A.S.Schwarz and Yu.S.Tyupkin, {\sl Phys. Lett.}
{\bf 59B}, 85 (1975).
}
In Euclidean region, Ashtekar's formalism is free from the
reality condition and may make a breakthrough in quantum theory of
gravity.\Ref\halfflat{For recent progress by Ashtekar,
	A.Ashtekar and J.Lewandowski, {\sl Projective techniques and
	functional integration for gauge theories, to appear in
	the special JMP issue on Functional Integration},
	ed. by C.DeWitt-Morette.
}

The purpose of the present paper is to give various four-dimensional
gravitational instanton solutions in Ashtekar's formalism.
This article is the detailed and enriched explanation of our previous
report.\Ref\IKF{
	H.Ishihara, H.Kubotani and T.Fukuyama,
	{\sl Proceedings of the sixth
	Marcel Grossmann Meeting on General Relativity } p.576
	ed. by H.Sato and T.Nakamura (World Scientific 1991).}
Explicitly, we examine Euclidean Bianchi type IX space because it is
especially suitable to treat in Ashtekar's formalism.\Ref\kodama{
	H.Kodama, {\sl Prog. Theor. Phys.} {\bf 80}, 1024 (1988).
}
We have not found the new solution of instanton.
However, the procedures in solving
Einstein equation are quite different from the conventional ones.
That is, instead of solving the self-duality condition, we fully
solve the constraints on the \lq initial surface\rq\ and
\lq Hamilton equations\rq .
So this formalism is applicable to the matter coupled systems with cosmological
constant.

This paper is organized as follows. In section two we give the general
framework of Ashtekar's formalism in homogeneous space with Euclidean
signature, which is applied to the special case of Bianchi type IX
space.
In section three we consider the most simple solutions of Bianchi type IX.
They give the vanishing $SO(3)$ field strength. It is shown that they
correspond to Eguchi-Hanson\Ref\EH{
 T.Eguchi and A.J.Hanson, {\sl Phys. Lett.} {\bf 74B}, 249 (1978).
}
and Euclidean Taub-NUT\Ref\TN{
	S.Hawking, {\sl Phys. Lett.} {\bf A60}, 2752 (1977).
}
solutions.
Then we proceed to discuss the same Bianchi IX space
with cosmological constant in section four.
In this case the field strength
cannot be zero from Hamiltonian constraint. However, we can find the very
simple solutions in the special cases. They are Euclidean
de Sitter and Fubini-Study solutions.\refmark\EGH
Furthermore, by adopting the general ansatz in field strength,
we can derive the Taub-NUT-de Sitter metric.\refmark\EH
Section five is devoted to
discussions. Througout of this paper we only consider four dimensional
Euclidean Einstein equations of homogeneous universes.

\chapter{General Framework}

Euclidean homogeneous universe is characterized by the line element
$$
  ds^2 = N^2 d\tau ^2 + \delta _{ab}~e_i^{~a} ~e_j^{~b} ~\s i ~\s j .
\eqn\ds
$$
Here $e_i^a$ are triad depending only on Euclidean time $\tau$.
First alphabets $a,b,\dots$ (middle alphabets $i,j,\dots$ )
indicate spatial part of internal indices
( that of world indices).
The symbols $\sigma^i$ are left invariant one-forms
which satisfy
$$
	d\sigma^i = C^i_{~jk} ~\s j \wedge ~\s k ,
\eqn\SO
$$
where $C^i_{~jk}$ is the structure constant.

Ashtekar's $SO(3)$ field strength is defined by
$$
  F^a = d A^a - {\kappa\over 2}\epsilon^a_{~bc} A^b\wedge A^c ,
\eqn\Fij
$$
where  $A^a$ is one form $A^a \equiv A^a_i\sigma^i$ and
$\kappa = 8\pi G $.
Eq.\Fij\ comes from the self-dual Riemann curvature,
$$
	{}^{+)}R^{0a} = d \w{}{0a}
				- \epsilon^a_{~bc} ~\w {}{0b}\wedge ~\w {}{0c}
\eqn\Riemann
$$
with an identification
$$
	A^a
		\equiv {2\over\kappa} ~\w {}{0a}
		\equiv{2\over\kappa}  (\omega^{0a}
			+ {1\over 2}\epsilon^a_{~bc} \omega^{bc}) .
\eqn\Aa
$$
For later convenience we will consider the action with the
cosmological constant.
The Jacobson-Smolin's action\Ref\js{
	 T.Jacobson and L.Smolin, {\sl Class. Quant. Grav.} {\bf 5},
	583 (1988).
}
$$
	S = {1\over\kappa}\int ~ ({}^{+)}R - \Lambda ) \sqrt{g} d^4x
\eqn\JS
$$
is transformed to
$$
  S = \int d^4 x \left[\e ia \dot\A ia + D_i\e ia\A 0a
	+ N^j \e ia \F {ji}a
	+ {1\over 2} \un \{ \epsilon_a^{~bc}\e ib \e jc \F {ij}a
	- {2\over\kappa} {\rm det}(\e ia )\Lambda \}\right]
\eqn\action
$$
by the (3+1) decomposition;
$$
 e^{00}=1/N , \quad  e^{i0} = N^i /N , \quad  e^{0a} = 0 , \quad e^{ia} =
{}^{(3)}e^{ia} .
\eqn\decomp
$$
Here
$$
	\un = N/{\rm det}(e_i^{~a}), \quad{\rm and}\quad
	\e ia = {\rm det}(e_i^{~a}) e^i_{~a} .
$$
Dot denotes derivative with respect to $\tau$ .
{}From Eq.\action\ we know that $\e ia$ and $\A ia$ are
canonical partners
and $\A 0a $ are multipliers for the internal $SO(3)$ gauge rotation.
Thus, we assume $\A ia$ are depending only on $\tau$ as $\e ia$.
Therefore, eq. \Fij\ is expressed as
$$
  \F {ij}a = 2 \A ka C^k_{~ij} - \kappa\epsilon^a_{~bc}\A ib \A jc .
\eqn\Fa
$$
The whole first class constraints are expressed as polynomial forms:
$$
\eqalignno{
  &\hbox{Hamiltonian\ constraint:} \quad
	\epsilon_a^{~bc} \F {ij}a \e ib \e jc
	 -{2\over \kappa}{\rm det}(\e ia )\Lambda = 0 , &\eqnalign{\HC} \cr
  &\hbox{Momentum\ constraint:} \quad
	~\F {ij}a \e ja = 0 ,			&\eqnalign\MC \cr
  &\hbox{Gauss'\ law\ constraint:} \quad
	~~D_i \e ia
	 = 2C^j_{~ji} \e ia - \kappa\epsilon_{ab}^{~~c}\A ib \e ic
	 = 0 .					&\eqnalign\GC \cr
}
$$

Time developements of new variables are given from the Poisson
bracket with Hamiltonian;
$$
\eqalignno{
	\dot{\tilde e}^{\hbox{\vbox{\vskip 3pt \hbox{$\scriptstyle i$}
				    \vskip -3pt}}}_{~a} &=
		\un (
		   C^i_{~jk} \epsilon_a^{~bc} \e jb \e kc
		- \kappa\e ia \e jb \A jb + \kappa\e ib \A jb \e ja ) ,
&\eqnalign\dyne\cr
	{\dot\A ia} &=
		-  \un ( \F{ij}{b} \epsilon_{b}^{~ac} \e jc
			- {1\over\kappa}\tilde \Delta_i^a \Lambda ),
&\eqnalign\dyna
}
$$
where  $\tilde \Delta _i^{~a}$ is $(i,a)$ element of the cofactor of
det($\e ia $).

In the subsequent sections we will deal with the
Bianchi type IX space. That is
$$
	C^i_{~jk} = \epsilon^i_{~jk} ,
\eqn\IX
$$
where $\epsilon^i_{~jk}$ is fully antisymmetric tensor
with $\epsilon^1_{~23}=1$.
In this case Ashtekar's formulation takes especially a simple form
and is solved explicitly.
Hereafter, we take $\kappa=2$ unit.

\chapter{Solutions without cosmological sonstant}

In this and the subsequent sections, we consider exact Euclidean
solutions of Einstein equation in the Bianchi type IX space.
In this section we consider the case with $\Lambda = 0$.

The system has twofold $SO(3)$ invariance.
One comes from the Ashtekar's field strength, that is, the
invariance with respect to the internal coordinates.
Another is due to the spatial property of Bianchi type IX universe,
i.e., the invariance with respect to the world coordinates.
These invariances allow twofold rotations
$$
	T^{~a}_b ~\e ia ~ {T'}^{~j}_i = \tilde {e'}^j_b ,
$$
then we can always set
$$
	\e ia = \diag (X(\tau ),Y(\tau ),Z(\tau )) .
\eqn\ediag
$$

The most simple solution of Eqs. \HC\ and \MC\ is the pure gauge solution
$$
	\vec F _{ij} \equiv \F{ij}a \tau_a = 0 ,
\eqn\pure
$$
where $\tau_a $ is the generator of $SO(3)$.
This solution is not trivial in contrast to the Yang-Mills
theory and does not imply a flat metric.
Hereafter, we recognize $i=1,2,3$ as $x,y,z$, respectively.
{}From Eq.\Fij\ , Eq.\pure\ takes the form
$$
\eqalign{
 &\hskip 2cm {1\over 2}\vec F_{xy}
		= \vec A_z - \vec A_x \times \vec A_y =0 \cr
 &{\hbox{ (cyclic permutation with respect to $x,y$\ and $z$)}} ,
}
\eqn\Fpure
$$
where $\vec A_x = A_x^{~a} \tau_a$ and the symbol $\times$ means
vector product in the internal space.
The solutions to Eq.\Fpure\ are
$$
\eqalignno{
	{\rm (i)}\quad &\vec A_x = \vec A_y = \vec A_z = 0
&\eqnalign\solone \cr
\noalign{or}
	{\rm (ii)}\quad &\vec A_x , \vec A_y , \vec A_z
	\hbox{~are orthonormal} .
&\eqnalign\soltwo
}
$$
The Gauss' law constraint in the Bianchi type IX space is reduced to
$$
	{\mib A}_i \times {\tilde{\mib e}}^i = 0
$$
Namely,  ${\mib A}_i$ is parallel to
$\tilde{\mib e}^i$ in the case (ii).
%
%

Equations of motion for $\A ia$ are automatically satisfied due to
the vanishing field strength. So we may concentrate on
the dynamical equations of $\e ia $ .
We will discuss the solutions (i) and (ii) separately in the
remaining part of this section.

\vskip 1cm
\noindent
{\bf 3-(i)} Vanishing $\vec A_i $

Substituting the diagonal form \ediag\ into the equation of
motion \dyne , we get
$$
	{\dot X}= 2 \un YZ,\quad
	{\dot Y}= 2 \un ZX,\quad
	{\dot Z}= 2 \un XY .
\eqn\dyn
$$
In order to solve Eq.\dyn\ we fix lapse function $\un$ as
$$
	\un = {1\over 4XYZ} .
\eqn\laps
$$
So ${\dot X}$ satisfies $X{\dot X}= {1/2}$ etc. and therefore
$$
	X = \sqrt{\tau-C_1},\quad Y = \sqrt{\tau-C_2}, \quad
	Z = \sqrt{\tau-C_3} ,
\eqn\sols
$$
where $C_1, C_2 {\rm and} C_3 $ are arbitraly constants.
The line element is
$$
	ds^2 = N^2 d\tau^2 + {YZ\over X}\sigma_x^{~2}
		+ {ZX\over Y}\sigma_y^{~2} + {XY\over Z}\sigma_z^{~2} .
\eqn\le
$$
Here the original lapse function $N$ is
$$
	N^2 \equiv \un ^2 ({\rm det}~e_i^{~a})^2 = ({1\over 4XYZ})^2 XYZ .
\eqn\olaps
$$
Thus from Eqs.\sols , \le\ and \olaps\ we obtain the metric
form given by Belinskii et al.\Ref\Bel{
	V.A.Belinskii, G.W.Gibbons, D.N.Page and C.N.Pope,
	{\sl Phys. Lett.} {\bf 76B}, 433 (1978).
}
$$
\eqalign{
  ds^2 = &{d\tau^2 \over 16 \sqrt{(\tau -C_1) (\tau -C_2) (\tau -C_3)}}
   + \sqrt{{(\tau -C_2)(\tau -C_3)\over (\tau -C_1)}}\sigma_x^{~2} \cr
\noalign{\vskip 5mm}
   &+ \sqrt{{ (\tau -C_3) (\tau -C_1)\over (\tau -C_2)}} \sigma_y^{~2}
   + \sqrt{{ (\tau -C_1) (\tau -C_2)\over (\tau -C_3)}} \sigma_z^{~2} .
}
\eqn\solmetric
$$
Let us introduce new coordinate $r$ defined by $r^4 =\tau $,
and assume axial symmetry, i.e., $C_1 = C_2 = a^4, C_3 = 0$.
Then we get
$$
	ds^2 = \{ 1-({a\over r})^4 \}^{-1} dr^2
    		+ r^2 \biggl[ \{ \sigma_x^{~2} + \sigma_y^{~2} \}
		+ \{ 1-({a\over r})^4 \}\sigma_z^{~2} \biggr] .
\eqn\eq
$$
This is nothing but Eguchi-Hanson metric.\refmark\EH

\vskip 1cm
\noindent
{\bf 3-(ii)} Orthonormal ${\vec A}_i$

In the case of orthonormal solution,
the discussions of the previous subsection
are valid with the modification of Eq.\dyn\ to
$$
\eqalign{
	{\dot X} &= 2 \un \{ YZ - X(Y+Z)\} ,\cr
	{\dot Y} &= 2 \un \{ ZX - Y(Z+X)\} ,\cr
	{\dot Z} &= 2 \un \{ XY - Z(X+Y)\} . \cr
}
\eqn\dynii
$$
We assume axial symmetry $X=Y$. Then Eq.\dynii\ is reduced to
$$
	{\dot X}= -2\un X^2 ,\quad
	{\dot Z}=  2\un \{ X^2- 2ZX\} .
\eqn\reducedeq
$$
If we fix lapse function $\un$ as
$$
	\un = {1\over 2X} ,
$$
Eq.\reducedeq\ is integrated to
$$
	X = C e^{-\tau}, \quad Z = D e^{-2\tau} + C e^{-\tau} ,
\eqn\solii
$$
where $C$ and $D$ are integration constants.
The original laps is
$$
	N^2 = \un^2 X^2Z = {Z\over 4} .
\eqn\laps
$$
Eqs.\le , \solii\ and \laps\ give the line element
$$
	ds^2 = {Z\over 4} d\tau^2 + Z (\sigma_x^{~2}+ \sigma_y^{~2})
				+ {X^2\over Z}\sigma_z^{~2} .
\eqn\le
$$
When $D=0$ the metric is flat.
When $D \neq 0$ we introduce new coordinate $r$ as
$$
	Z = D e^{-2\tau} + C e^{-\tau} = \hbox{sgn}(D)(r^2 - M^2) ,
\eqn\newcoordinate
$$
where sgn$(D)$ is the signature of $D$ and $M^2 = C^2/4\vert D\vert$.
Then $X$ becomes
$$
	X= 2\hbox{sgn}(D)M(\pm r-M)
$$
from Eqs.\solii\ and \newcoordinate .
Thus we obtain the line element
$$
	ds^2=\pm\left[{1\over 4}{r+ M\over r - M} dr^2
		+ (r^2-M^2) \{\sigma_x^{~2} + \sigma_y^{~2} \}
		+ 4M^2 {r - M\over r + M} \sigma_z^{~2}\right] .
\eqn\eq
$$
This is the Euclidean Taub-NUT metric.

\chapter{ Solution with cosmological constant}

Anti self-duality has solved the \lq Euclidean\rq\ initial value
problems
and we have obtained Eguchi-Hanson (E-H)
and Euclidean Taub-NUT (T-N) solutions. As was indecated by
Eguchi and Hanson,\refmark\EH
E-H and T-N solutions compose the triplet together
with Fubini-Study (F-S) solution.
The last one is the solution to the Einstein gravity with cosmological
constant. The presence of cosmological constant prevents
the anti duality to be the solution to Hamiltonian constraint.
In this section we consider
gravitational instantons with cosmological constant known as
Taub-NUT-de Sitter solution.

Hamiltonian constraint \HC\ suggests that $\F{ij}a$ is related by
$\e ia$ algebraically. So we assume
$$
	\F{ij}a = M^{ab}\e kb \epsilon_{kij} .
\eqn\ansatz
$$
{}From momentum constraint \MC , $M^{ab}$ is symmetric.
For diagonal $\e ia$ ,
$$
 \e ia = {\rm diag} (X(\tau ), Y(\tau ), Z(\tau ) ) ,
\eqn\ediag
$$
Hamiltonian constraint leads
$$
	M^a_a ={\Lambda\over 2} .
\eqn\tr
$$

{}From the equation of motion \dyne , it is obvious that
the diagonal form of $\A ia$,
$$
 \A ia = {\rm diag} (\alpha (\tau ), \beta (\tau ), \gamma (\tau )) ,
\eqn\adiag
$$
guarantees $\e ia$ to evolve retaining its diagonal form.
Eqs.\ediag\ and \adiag\ satisfy momentum and Gauss' law constraint
trivially in the Bianchi IX space.
For the diagonal  $\A ia$ of Eq.\adiag , non-vanishing components of
$SO(3)$ field strength become
$$
\eqalign{
	\F{xy}3 = 2(\gamma -\alpha\beta ), \cr
	\F{yz}1 = 2(\alpha -\beta\gamma ), \cr
	\F{zx}2 = 2(\beta -\gamma\alpha ). \cr
}
\eqn\rF
$$
Thus, $M^{ab}$ has only diagonal components. Then Eq.\ansatz\ is reduced to
$$
	\F{ij}a = \mu^a \e k a \epsilon_{kij} \quad
	(a: \hbox{no summation}) ,
\eqn\dm
$$
with
$$
	\sum_a \mu^a = {\Lambda\over 2} .
\eqn\tr
$$
The ansatz Eq.\dm\ with Eq.\tr\ is an extension of the Ashtekar-Renteln
or the Samuel ansatz.\Ref\ARS{
	A.Ashtekar and P.Renteln, unpublised (1987), \nextline
	J.Samuel, {\sl Class. Quantum Grav.} {\bf 5}, L123 (1988), \nextline
	S.Koshti and N.Dadhich, {\sl Class. Quantum Grav.} {\bf 5}, L5 (1990) .
}

Here we, furthermore, assume axial symmetry $X=Y$.
The invariant line element reads
$$
	ds^2= N^2 d\tau^2 + Z(\sigma_x^{~2} + \sigma_y^{~2})
		+{X^2\over Z}\sigma_z^{~2} ,
\eqn\lineelem
$$
where the laps function is given by
$$
	N^2=\un^2 X^2 Z .
\eqn\laps
$$

The equation of motion \dyne\ leads us to $\alpha = \beta$, then
$$
\eqalign{
	\dot X &= 2 \un X \{ Z (1-\gamma ) - \alpha X \}, \cr
	\dot Z &= 2 \un X ( X -2 \alpha Z ). \cr
}
\eqn\rdyne
$$
Similarly, Eq.\dyna\ requires
$$
\eqalign{
	\dot\alpha &= 2 \un \alpha ( 1- \gamma ) Z , \cr
	\dot\gamma &= 2 \un {X^2\over Z} (\gamma - \alpha^2 ) . \cr
}
\eqn\rdyna
$$
In Eq.\rdyna\ use has been made of Eq.\HC .

{}From Eqs. \rF\ and \dm , we obtain
$$
\eqalign{
	&2(\gamma-\alpha^2) = \mu^3 Z , \cr
	&2\alpha(1-\gamma) = \mu^1 X  \cr
}
\eqn\redan
$$
with $2\mu^1 + \mu^3 = {\Lambda / 2}$.

Compatibility of the equations of motion for $X, Z, \alpha$ and
$\beta$ gives the evolution equations for $\mu^a$
$$
\eqalign{
	\dot\mu^1 &= -2\un \alpha X(\mu^3 -\mu^1), \cr
	\dot\mu^3 &= 4\un \alpha X(\mu^3 -\mu^1) .
}
\eqn\eqmu
$$
Since non-degenerate metric requires $X\neq 0$ then $\mu^a$ can be constant
when $\alpha=0$ or $\mu^1=\mu^3$. Otherwise, Eq.\eqmu\ forces $\mu^a$
to be also functions of $\tau$.
We discuss these cases separately.

\vskip 1cm
\noindent
{\bf 4-(i)} $\alpha =0$ ($\mu^1=0$) case

In $\alpha =0$ case, from Eqs.\tr\ and \redan , we get
$\mu^1=0$, $\mu^3={\Lambda/2}$ then
$$
	4\gamma = \Lambda Z .
\eqn\rhamil
$$

We find the equations of motion for $X, Z$ from Eqs.\rdyne\ and \rhamil ,
$$
\eqalign{
	\dot X &= 2 \un XZ (1-{\Lambda\over 4}Z ) , \cr
	\dot Z &= 2 \un X^2  .
}
\eqn\eq
$$
By taking the laps function as
$$
	\un = {1\over 2X^2} ,
\eqn\gauge
$$
Eq.\eq\ is easily solved as follows
$$
\eqalign{
	&Z = \tau - \tau_0 , \cr
	&X^2 = -{\Lambda\over 6}(\tau - \tau_0)^3 + (\tau - \tau_0)^2 + C ,
}
\eqn\sol
$$
where $\tau_0$ and $C$ are integration constants.
Using Eq.\rhamil\ we get
$$
	\alpha =0, \quad \gamma = {\Lambda\over 4}(\tau - \tau_0) .
\eqn\solag
$$
{}From Eq.\lineelem\ with Eqs.\laps\ and \gauge , we obtain the metric
in the form
$$
	ds^2= {Z\over 4X^2} d\tau^2 + Z(\sigma_x^{~2} + \sigma_y^{~2})
		+{X^2\over Z}\sigma_z^{~2} ,
\eqn\metric
$$
where $X, Z$ are given by Eq.\sol .
In $\Lambda = 0$ limit, Eq.\solag\ shows $\A ia =0$. Therefore, this solution
is
an extended one of E-H solution to $\Lambda \ne 0$ case.

Let us introduce $r$ instead of $\tau$ by
$$
	Z =\tau-\tau_0 = r^2 ,
$$
then we get
$$
	ds^2 = \{ 1-({a\over r})^4 -{\Lambda\over 6}r^2 \}^{-1} dr^2
    		+ r^2 \biggl[ \{ \sigma_x^{~2} + \sigma_y^{~2} \}
  	 + \{ 1-({a\over r})^4-{\Lambda\over 6}r^2 \}\sigma_z^{~2} \biggr] ,
\eqn\exEH
$$
where we have set $C = -a^4$ .

When $C=0$ the metric \metric\ is Fubini-Study metric.
To see this explicitely, we introduce another coordinate $\rho$ by
$$
	Z= {\rho^2\over 1+{\Lambda\over 6} \rho^2}
\eqn\eq
$$
and set $C=0$. Then it follows that
$$
	X^2= {\rho^4\over ({1+{\Lambda\over 6} \rho^2})^3} .
\eqn\eq
$$

Temporal part is
$$
	N^2d\tau^2 = \un^2 X^2Z d\tau^2
		= {d\rho^2\over (1+{\Lambda\over 6} \rho^2)^2 }.
$$
Therefore, the invariant line element turns out to be
$$
	ds^2 = {1\over({ 1+{\Lambda\over 6} \rho^2 })^2}( d\rho^2
		+ \rho^2 \sigma_z^{~2})
	+ {\rho^2\over 1+{\Lambda\over 6} \rho^2 }
			(\sigma_x^{~2} + \sigma_y^{~2})
\eqn\FS
$$
from Eq.\lineelem . This is the Fubini-Study metric in conventional form.

\vskip 1cm
\noindent
{\bf 4-(ii)} $\mu^1=\mu^3$ case

In this case, $\mu^a={\Lambda\over 6}$ then \ansatz\ becomes\refmark\IKF
$$
	\F{ij}a = {\Lambda\over 3!}\epsilon_{ijk} \tilde e^{ka} .
\eqn\newansatz
$$
This case is discussed by some authors.\refmark\ARS
Equation \newansatz\ together with Eq.\rF\ gives
$$
\eqalign{
	&\gamma-\alpha^2 = {\Lambda\over 12} Z , \cr
	&\alpha(1-\gamma) = {\Lambda\over 12} X .
}
\eqn\rhc
$$
Equation of motions \rdyne\ and \rdyna\ under ansatz \rhc\ are
$$
\eqalign{
	\dot X &= 2 \un X \{ Z (1-\gamma ) - \alpha X \}, \cr
	\dot Z &= 2 \un X ( X -2 \alpha Z ), \cr
}
\eqn\eomx
$$
and
$$
\eqalign{
	\dot\alpha &= \un{\Lambda\over 6} XZ , \cr
	\dot\gamma &= \un{\Lambda\over 6} X^2 . \cr
}
\eqn\eoma
$$
So, fixing the gauge of $\un$ as
$$
	\un = {1\over 2X^2} ,
\eqn\gauge
$$
we can solve Eq.\eoma . First, $\gamma$ is immediately
given by
$$
	\gamma = {\Lambda\over 12}( \tau - \tau_0 ).
\eqn\solg
$$
As for $\alpha$ , $\dot\alpha$ is reduced to
$$
	\dot\alpha = {\Lambda\over 12} {\gamma-\alpha^2\over \alpha(1-\gamma)}
\eqn\da
$$
by use of Eqs.\eoma\ and \rhc .
Substituting Eq.\solg\ into Eq.\da\ we obtain the solution $\alpha$
in the form
$$
	\alpha^2 = 2 \Big({\Lambda\over 12} (\tau - \tau_0)-1 \Big) +1
	+ {12\over\Lambda}C \Big({\Lambda\over 12}(\tau - \tau_0)-1 \Big)^2 ,
\eqn\sola
$$
where $C$ is an arbitrary constant.
We can set $\tau_0$ as ${\Lambda\over 12}\tau_0 = -1$
without loss of generality.
Then
$$
\eqalign{
	&\alpha^2 = 1 + {\Lambda\over 12} (2\tau + C \tau^2) , \cr
	&\gamma = 1 + {\Lambda\over 12} \tau .
}
\eqn\solA
$$
In $\Lambda=0$ limit, $\alpha=\gamma=1$. Thus the solution for 4-(ii)
is an extended one of the Taub-NUT
solution to $\Lambda \neq 0$ case.

Without solving Eq.\eomx, we get $X,Z$ immediately
from Eqs.\solA\ and \rhc\ in the form
$$
\eqalign{
	&Z=-( \tau + C \tau^2 ) \cr
	&X^2=\Big\{ 1 + {\Lambda\over 12}( 2\tau
		+ C \tau^2) \Big\}\tau^2 .
}
\eqn\sol
$$
{}From Eq.\lineelem\ with Eqs.\laps\ and \gauge , we obtain
$$
	ds^2= {Z\over 4X^2} d\tau^2 + Z(\sigma_x^{~2} + \sigma_y^{~2})
		+{X^2\over Z}\sigma_z^{~2} ,
\eqn\metric
$$
where $X,Z$ are given by Eq.\sol.

We introduce new coordinate $\rho$ instead of $\tau$ by
$$
	Z=-(\tau+ C\tau^2)=\cases{
		\rho^2 		&\hbox{for $C=0$} \cr
		\hbox{sgn}(C)(L^2 - \rho^2 )	&\hbox{for $C\neq 0$} ,
		}
$$
where $\displaystyle L^2 \equiv {1\over 4\vert C\vert}$. Then
$$
	X^2=\cases{
		\rho^4(1-{\Lambda\over 6}\rho^2)
	 		&\hbox{for $C=0$} \cr
		4L^2\{(\rho-L)^2
		  +\hbox{sgn}(C)
		{\Lambda\over 12}(-3L^4+8L^3\rho-6L^2\rho^2 + \rho^4)\}
			&\hbox{for $C\neq 0$} ,
		}
$$
and
$$
	N^2d\tau^2=\cases{
		(1-{\Lambda\over 6}\rho^2)^{-1}d\rho^2
	 		&\hbox{for $C=0$} \cr
 	{\displaystyle
		L^2 {\hbox{sgn}(C) ( L^2-\rho^2)\over X^2} d\rho^2 }
			&\hbox{for $C\neq 0$} \cr
		}
$$
In the case of $C=0$, the metric is Fubini-Study in the form
of Eq.\exEH\ with $a=0$.

On the other hand, when $C\neq 0$, the result is a special case of
the Taub-NUT-de Sitter metric. This fact will be presented in the next
subsection.

Eq.\sol\ requires that the metric becomes isotropic when $C=\Lambda/12$.
It is nothing but the de Ditter ($C >0$) or anti-de Sitter
($C<0$) solution.

\vskip 1cm
\noindent
{\bf 4-(iii)} $\mu^a\neq const.$ case

Fixing the gauge of $\un$ as
$$
	\un = {1\over \alpha X}
\eqn\gauge
$$
we can solve Eq.\eqmu\ in the form
$$
\eqalign{
	\mu^1 &= {\Lambda\over 6}\{\mu_0 e^{6\tau} + 1\},  \cr
	\mu^3 &= {\Lambda\over 6}\{-2\mu_0 e^{6\tau} + 1\},  \cr
}
\eqn\eq
$$
where $\mu_0$ is an arbitrary non-vanishing constant. Since the absolute
value of $\mu_0$ can be removed by a translation of $\tau$,
we may only note the signature of $\mu_0$.

Together with Eqs.\redan\ and \gauge , Eq.\rdyna\ requires
$$
\eqalign{
	&{1\over 2}(\alpha^2)\dot{}
		= 2{\mu^1\over \mu^3} (\gamma - \alpha^2),  \cr
	&\dot\gamma = 2 {\mu^3\over \mu^1} (1-\gamma) .
}
\eqn\eom
$$
Substituting the solutions of $\mu^1$ and $\mu^3$ into it, Eq.\eom\
is easily integrated to
$$
\eqalign{
	&\gamma={C\Lambda\over 12} \{e^{4\tau}\pm e^{-2\tau}\}+1 , \cr
	&\alpha^2={C\Lambda\over 12}  \{-e^{4\tau}\pm 2e^{-2\tau}\}+1
			+ {D\Lambda\over 12}  \{2e^{2\tau} \mp e^{-4\tau}\} .
}
\eqn\solalpha
$$
Here $C$ and $D$ are integral constants and the double sign corresponds
to the signature of $\mu_0$.
Thus having solved $\mu^a$ and $\A ia$, we can derive the explicit form of
$\e ia$ from Eq.\redan ,
$$
\eqalign{
	&X^2 = (Ce^{-2\tau})^2 \Big\{{C\Lambda\over 12}
		(-e^{4\tau}\pm 2e^{-2\tau}) +1
		+ {D\Lambda\over 12}\Lambda (2e^{2\tau} \mp e^{-4\tau})\Big\} , \cr
	&Z = \pm \{-C e^{-2\tau}+ D e^{-4\tau} \} .
}
\eqn\solX
$$
Eq.\lineelem\ together with Eqs.\laps\ and \gauge\ requires the
invariant line element to be
$$
	ds^2= {Z\over \alpha^2} d\tau^2 + Z(\sigma_x^{~2} + \sigma_y^{~2})
		+{X^2\over Z}\sigma_z^{~2} .
\eqn\metric
$$
$X,Z$ and $\alpha$ in Eq.\metric\ are given by Eqs.\solX\ and \solalpha .

We introduce new coordinate $\rho$ in place of $\tau$ by
$$
	Z=\pm \{-C e^{-2\tau}+ D e^{-4\tau}\} = \cases{
		\rho^2 &\hbox{ for $D=0$} \cr
		\pm\hbox{sgn}(D) (\rho^2 - L^2)
			&\hbox{ for $D\neq 0$}
		}
$$
where $\displaystyle L^2\equiv {C^2\over 4\vert D\vert}$ .
The range of $\rho$ should be chosen so that $Z$ is positive.
Thus,
$$
	N^2 d\tau^2 = {Z\over\alpha^2}d\tau^2
		= \cases{
		\{ 1 - {C\Lambda\over 12}{C^2\over\rho^4}
			- {\Lambda\over 6}\rho^2 \}^{-1} d\rho^2
			&\hbox{for $D=0$} \cr
		{\displaystyle
		 \pm L^2 {\hbox{sgn}(D) (\rho^2 - L^2)\over X^2} d\rho^2 }
			&\hbox{for $D\neq 0$} \cr
		}
\eqn\temporal
$$
and
$$
	X^2 = \cases{
		\rho^4\{ 1 - {C\Lambda\over 12}{C^2\over\rho^4}
				- {\Lambda\over 6}\rho^2 \}
			&\hbox{for $D=0$} \cr
	\eqalign{
		4L^2\Big[\rho^2&-(2+{C^3\over 4L^2})\rho+L^2  \cr
		&+\hbox{sgn}(D){\Lambda\over 4}(L^4+{2\over 3}L^3\rho
				+2L^2\rho^2 -{1\over 3}\rho^4)\Big]
	}
			&\hbox{for $D\neq 0$} \cr
		}
\eqn\solX
$$
In the case of $D=0$, the line element becomes
the extended E-H metric Eq.\exEH .
On the other hand, in the case of  $D\neq 0$, we obtain
$$
	ds^2=\pm\left[{\rho^2-L^2\over 4\Delta} d\rho^2
		+ (\rho^2-L^2)(\sigma_x^{~2} + \sigma_y^{~2})
		+{4L^2\Delta\over \rho^2-L^2} \sigma_z^{~2}\right] .
\eqn\TN
$$
Here $\Delta$ is defined by
$$
	\Delta \equiv \rho^2-2M\rho+L^2
		\pm{\Lambda\over 4}(L^4+2L^2\rho^2 -{1\over 3}\rho^4)
$$
with
$$
	M \equiv {1\over 3 }\Lambda L^2 + 1 + {C^3\over 8L^2} .
$$
This is nothing but the Taub-NUT-de Sitter metric.

\chapter{Discussion}

In this article we have discussed about the explicit solutions
of Euclidean Einstein equation of Bianchi type IX universe.
The equations we have treated are the non linear ordinary differential
equations with many dependent variables.
As we have mentioned the non-linearity of the equations are
drastically reduced in the Ashtekar formalism.

Further developement beyond this article we may look in two routes.
One is to study the homogeneous space other than Bianchi type IX
universe. In Bianchi type IX case, invariance groups of the field
strength and of the considering space are both $SO(3)$.
This coincidence brings about the additional simplicity.
However, the loss of this simplicity does not change the fundamental
framework mentioned above and our formulation seems to be applicable
to the various types of homogeneous universes.
Developement in another route may be more substantial,
which is to extend independent
variables to two, three and finally four. (As a direct extension we may
consider multi instanton extension of the solutios obtained in this
article.\Ref\NJH{
	N.J.Hitchin,
	{\sl Math. Proc. Camb. Phil. Soc.} {\bf 83}, 465 (1979), \nextline
	P.B.Kronheimer, {\sl J.Diff. Geometry} {\bf 29}, 665 (1989).
}
)
In this route we are confronted with the integrable non-linear sciences.
In this well established region we have already the very poweful tools
and concepts such as Lax pair, Painlev\'e properties, inverse scattering
mathod and Hirota's direct method and so on.\Ref\soliton{
	See, for instance, M.J.Ablowitz and P.A.Clarkson,
	{\sl Soliton, Nonlinear Evolution Equations and Inverse Scattering}
	(Cambridge University Press 1991).
}

It is very  intersting to ask what the Ashtekar formalism can add to this
fertile field.
\refout
\end